\documentclass[runningheads]{llncs}

\usepackage[english]{babel}
\usepackage{amsmath}

\begin{document}

\title{On Duplication in Mathematical Repositories
\thanks{The final publication of this paper is available at www.springerlink.com.}}
\titlerunning{On Duplication in Mathematical Repositories}

\author{Adam Grabowski\inst{1}
\and
Christoph Schwarzweller\inst{2}}

\institute{Institute of Mathematics, University of Bia{\l}ystok\\
   ul. Akademicka 2, 15-267 Bia{\l}ystok, Poland\\
            \email{adam@math.uwb.edu.pl}
\and
           Department of Computer Science, University of Gda\'{n}sk\\
           ul. Wita Stwosza 57, 80-952 Gda\'nsk, Poland\\
            \email{schwarzw@inf.ug.edu.pl}}

\maketitle

\begin{abstract}
Building a repository of proof-checked mathematical knowledge is
without any doubt a lot of work, and
besides the actual formalization process there also is the task of
maintaining the repository.
Thus it seems obvious to keep a repsoitory as small as possible,
in particular each piece of mathematical knowledge should be formalized only once.

In this paper, however, we claim that it might be reasonable or even necessary
to duplicate knowledge in a mathematical repository. 
We analyze different situations and reasons for doing so and 
provide a number of examples supporting our thesis.
\end{abstract}

\section{Introduction}


Mathematical knowledge management aims at providing both tools 
and infra\-structure supporting the organization, development,
and also teaching of mathematics using modern techniques
provided by computers.
Consequently, large repositories of mathematical knowledge are of
major interest because they provide users with a
data base of --- verified --- mathematical knowledge.
We emphasize the fact that a repository should contain
verified knowledge only together with the corresponding proofs.
We believe that (machine-checked or -checkable) proofs necessarily belong
to each theorem and therefore are an essential part of a
repository.

However, mathematical repositories should be more than
collections of theorems and their proofs accomplished by a 
prover or proof checker.
The overall goal here is not only stating and proving a theorem --- though this remains
an important and challenging part --- but also presenting definitions and 
theorems so that the ``natural'' mathematical buildup remains visible.
Theories and their interconnections should be available, so that the further
development of the repository can be based upon these.
Being not trivial as such,
this becomes even harder to assure for an open repository with a large number of
authors. 

In this paper we deal with yet another organizational aspect of building mathematical
repositories: the duplication of knowledge,
by which we mean that a repository includes redundant knowledge.
At first glance this may look inacceptable or at least unnecessary.
Why should one include --- and hence formalize --- the same thing more than once?
A closer inspection, however, shows that mathematical redundance may occur in different
non-trivial facets:
Different proofs of a theorem may exist or different versions of a theorem
formulated in a different context.
Sometimes we even have different representations of the same mathematical object
serving for different purposes.

From the mathematical point of view this is not only harmless but also desirable;
it is part of the mathematical progress that theorems and definitions change and evolve.
In mathematical repositories, however, each duplication of knowledge causes an additional amount
of work.
In this paper we analyze miscellanous situations and reasons why there could --- and
should --- be at least some redundance in mathematical repositories.
These situations range from the above mentioned duplication of proofs, theorems and representations
to the problem of generalizing knowledge.
Even techical reasons due to the progress of a repository may lead
to duplication of knowledge.

\section{Different Proofs of a Theorem} \label{secCRT}

The Chinese Remainder Theorem is a result about congruences over the integers.
It states that an integer $u$ can be completely described by the sequence
of its remainders --- if the number of remainders is big enough.
The ``standard'' version of the theorem reads as follows.

\begin{theorem} \label{cong} \rm 
Let $m_1, \, m_2, \, \ldots , m_r$ be positive integers such that $m_i$ and $m_j$
are relatively prime for $i \neq j$.
Let $m = m_1 m_2 \cdots m_r$ and let $u_1, \, u_2, \, \ldots , u_r$
be integers.
Then there exists exactly one integer $u$ with 
\[ 0 \leq u < m \ \mbox{ and } \ u \ \equiv \ u_i \mbox{ mod } m_i 
\ \mbox{ for all } \ 1 \leq i \leq r. \ \ \diamond \]
\end{theorem}

In the following we present three different proofs of the theorem
and discuss their relevance to be included in mathematical repositories.
It is very easy to show, that there exists at most one such integer $u$;
in the following proofs we therefore focus on proving the existence of $u$.
The proofs are taken from \cite{knuth}. \\

\noindent
{\bf First proof:} 
Suppose integer $u$ runs through the $m$ values $0 \leq u < m$.
Then $(u \mbox{ mod } m_1, \ldots , u \mbox{ mod } m_r)$ also runs through
$m$ different values, because the system of congruences has at most
one solution. 
Because there are exactly $m_1 m_2 \cdots m_r = m$ different tuples
$(v_1, \ldots , v_r)$ with $0 \leq v_i < m_i$, every tuple occurs exactly
once, and hence for one of those we have 
$(u \mbox{ mod } m_1, \ldots , u \mbox{ mod } m_r)$ $ = (u_1, \ldots , u_r)$.$ \ \diamond$ \\

This proof is pretty elegant and uses a rather obvious 
variant of the pigeon hole principle:
If we pack $m$ items without repetition to $m$ buckets, then we must have
exactly one item in each bucket.
It is therefore valuable to include this proof in a repository for didactic or
aesthetic reasons.
On the other hand, formalization of the proof is not necessarilly straightforward.
One has to argue about the number of different $r$-tuples and,
more importantly, to show that there exists a bijecton
between the set of $r$-tuples and the non-negative integers smaller than $m$.
Another disadvantage is that the proof is non-constructive, so that it
gives no hints to find the value of $u$ --- besides the rather valueless
``Try and check all possibilities, one will fit''. 
This is even more disturbing, because a constructive proof can easily be given:\\

\noindent
{\bf Second proof:}
We can find integers $M_i$ for $1 \leq i \leq r$ with
\[ M_i \equiv 1 \mbox{ mod } m_i \ \mbox{ and } \
   M_j \equiv 0 \mbox{ mod } m_i \ \mbox { for } j \neq i . \]
Because $m_i$ and $m/m_i$ are relatively prime, we can take for example
\[ M_i = (m/m_i)^{\varphi(m_i)}, \]
where $\varphi$ denotes the Euler function. 
Now, 
\[ u = (u_1M_1 + u_2M_2 + \cdots + u_rM_r) \mbox{ mod } m \]
has the desired properties. $\ \diamond$ \\

This proof constructs $r$ constants $M_i$ with which the sought-after $u$
can easily be computed. It therefore, in some sense, contains more
information than the first proof,  that should be contained 
in the repository also.
The proof uses far more evolved mathematical notations --- namely Euler's
function --- and for that reason may also be considered more interesting than the 
first one.
Formalization requires the use of Euler's function\footnote{Actually
a mild modification of the proof works without
Euler's function.} which may cause 
some preliminary work.
From a computer science point of view the proof has two disadvantages.
First, it is not easy to compute Euler's function;
in general one has to decompose the moduli
$m_i$ into their prime factors.
Second, the $M_i$ being multiples of $m/m_i$ are really big numbers, so that
a better method for computing $u$ is highly desirable.
Such a method has indeed been found by H. Garner, which gives a third proof
of Theorem 1: 
\\

\noindent
{\bf Third proof:}
Because we have $\gcd(m_i,m_j) = 1$ for $i \neq j$
we can find integers $c_{ij}$ for $1 \leq i < j \leq r$ with
\[ c_{ij} m_i \equiv 1 \mbox{ mod } m_j  \]
by applying the extended Euclidean algorithm to $m_i$ and $m_j$.
Now taking
\[\begin{array}{lcl}
v_1 & := & u_1 \mbox{ mod } m_1 \\
v_2 & := & (u_2 - v_1) c_{12} \mbox{ mod } m_2 \\
v_3 & := & ((u_3 - v_1)c_{13} - v_2) c_{23} \mbox{ mod } m_3 \\
& & \vdots \\
v_r & := & (\ldots ((u_r - v_1) c_{1r} - v_2) c_{2r} - \cdots - v_{r-1}) c_{(r-1)r}
       \mbox{ mod } m_r 
\end{array}\]
and then setting
\[ u := v_rm_{r-1} \cdots m_2m_1 + \cdots + v_3m_2m_1 + v_2m_1 + v_1 \]
we get the desired integer $u$. $\ \diamond$\\

The proof uses $\binom{r}{2}$ constants $c_{ij}$ that can be computed
with the extended Euclidean algorithm because we have $\gcd(m_i,m_j) = 1$ for
$i \neq j$.
When constructing the $v_i$ the application of the modulo operation in each step ensures
that the occurring values remain small.
The proof is far more technical than the others in constructing 
$\binom{r}{2}+r$ additional constants, the $v_i$ in addition being recursively
defined. 
On the other hand, however, this proof includes an efficient method to compute
the integer $u$ from Theorem 1.   

We see that the question which proof of a theorem should be formalized,
does not only depend on the hardness of the formalization in a given system.
Both elegance and the amount of information are issues that can be taken into
consideration --- this may even result in formalizing more than
one proof.

\section{Different versions of Theorems}


There are quite a number of reasons why different versions of the same
theorem exist and may be included in mathematical repositories.
Besides mathematical issues we also identified reasons justified
by formalization issues or the development of repositories itself.
For illustration we again use the CRT as an example.

\subsection{Restricted Versions}

Theorems are not always shown with a proof assistant to be included in
a repository in the first place:
Maybe the main goal is to illustrate or test a new implemented proof
technique or just to show that this special kind of mathematics can
be handled within the particular system.
In this case it is often sufficient --- or simply easier --- to prove a weaker or restricted
version of the original theorem from the literature.

In Hol Light \cite{hollight}, for example, we find the following theorem.

{\small\begin{verbatim}
# INTEGER_RULE
  '!a b u v:int. coprime(a,b) ==> 
                       ?x. (x == u) (mod a) /\ (x == v) (mod b)';
\end{verbatim}}

\noindent
This is a version of the Chinese Remainder Theorem \ref{cong}
stating that in case of two moduli {\tt a} and {\tt b} only there exists a simultaneous 
solution {\tt x} of the congruences.
Similar versions have been shown with {\tt hol98} (\cite{hurd}), the Coq proof
assistant (\cite{coq}) or Rewrite Rule Laboratory (\cite{zhang}).

From the viewpoint of mathematical repositories it is of course desirable
to have included the full version of the theorem also.
Can we, however, in this case easily set the restricted version aside?
Note that the above theorem in Hol Light also serves as a rule for proving
divisibility properties of the integers.
Erasing the restricted version then means that the full version has to be
used instead.
It is hardly foreseeable whether this will work for all other proofs relying on 
the restricted version.
So, probably sometimes both the restricted and the full version belong to the 
repsository.

\subsection{Different Mathematical Versions}

The most natural reason for different versions of theorems is that mathematicians
often look at the same issue from different perspectives.
The CRT presented in Section \ref{secCRT} deals with 
congruences over the integers: it states the existence of an integer
solving a given system of congruences.
Looking from a more algebraic point of view 
we see that the moduli $m_i$
can be interpreted as describing the residue class rings $\mathcal{Z}_{m_i}$.
The existence and uniqueness of the integer $u$ from the CRT
then gives rise to an isomorphism between rings \cite{modern}:

\begin{theorem} \rm \label{alg}
Let $m_1, \, m_2, \, \ldots , m_r$ be positive integers such that $m_i$ and $m_j$
are relatively prime for $i \neq j$ and
let $m = m_1 \, m_2 \, \cdots m_r$.
Then we have the ring isomorphism
\[ \mathcal{Z}_m \ \cong \  
\mathcal{Z}_{m_0} \times \cdots \times  \mathcal{Z}_{m_r}. \ \ \diamond \]
\end{theorem}

This version of the CRT has been formalized in {\tt hol98} \cite{hurd}.
Here we find a two-moduli version that in addition is restricted
to multiplicative groups.
Technically, the theorem states that for relative prime moduli $p$ and $q$ the function
$\lambda x. (x\ \mbox{mod}\ p,x\ \mbox{mod}\ q)$ is a group isomorphism
between $\mathcal{Z}_{pq}$ and $\mathcal{Z}_{p} \times \mathcal{Z}_{q}$.

\[\begin{array}{l}
\vdash \forall p,q. \\
\hspace*{0,5cm}1 < p \wedge 1 < q \wedge {\bf gcd}\ p\ q = 1 \Rightarrow \\
\hspace*{0,5cm}(\lambda x. (x\ \mbox{mod}\ p,x\ \mbox{mod}\ q)) \in \\
\hspace*{0,85cm}{\bf group\_iso}\ ({\bf mult\_group}\ pq) \\
\hspace*{0,85cm}({\bf prod\_group}\ ({\bf mult\_group}\ p)\ ({\bf mult\_group}\ q))
\end{array}\]

\noindent
Note that, in contrast to Theorem \ref{alg}, the isomorphism is part 
of the theorem itself and not hidden in the proof.

It is not easy to decide which version of the CRT
may be better suited for inclusion in a mathematical repository.
Theorem \ref{alg} looks more elegant and in some sense contains more information
than Theorem \ref{cong}: It does not state the existence of a special integer,
but the equality of two mathematical structures.
The proof of Theorem \ref{alg} uses the homomorphism theorem for rings and is 
therefore interesting for didactic reasons, too.
On the other hand, Theorem \ref{cong} uses integers and congruences
only, so that one needs less preliminaries to understand it.
Theorem \ref{cong} and its proof also give more information than theorem \ref{alg}
concerning computational issues\footnote{To apply the homomorphism theorem in the proof of Theorem
\ref{alg} one needs to show that the canonical homomorphism is a surjection
with kernel $(m)$. This sometimes is done by employing the extended 
Euclidean algorithm, so that this proof gives an algorithm, too.}
 --- at least if not the first proof only has been formalized.

\subsection{Different Technical Versions}

Another reason for different versions of a theorem may be originated in the
mathematical repository itself.
Here again open repositories play an important role:
Different authors, hence different styles of formalizing and different kinds of mathematical
understanding and preferences meet in one repository.
So, it may happen that two authors formalize the same (mathematical) theorem,
but choose a different formulation and/or a different proof.
We call this technical versions. 

Especially in evolving systems such versions may radically differ
just because the system's language improved over the years.
In the Mizar Mathematical Library, for example, 
we find the following CRT \cite{int6}

{\small\begin{verbatim}
theorem
for u being integer-yielding FinSequence,
    m being CR_Sequence st len u = len m
ex z being Integer
st 0 <= z & z < Product(m) & for i being natural number 
   st i in dom u holds z,u.i are_congruent_mod m.i;
\end{verbatim}}

Here, a \verb@CR_Sequence@ is a sequence og natural numbers, which are pairwise
relative prime.
Note that the formulation f the CRT is very close to the texbook version theorem \ref{cong}.

In another Mizar article \cite{wsierp1}, however, we find a different formulation of the CRT:

{\small\begin{verbatim}
theorem :: WSIERP_1:44
len fp>=2 &
(for b,c st b in dom fp & c in dom fp & b<>c holds (fp.b gcd fp.c)=1)
implies for fr st len fr=len fp holds ex fr1 st (len fr1=len fp &
for b st b in dom fp holds (fp.b)*(fr1.b)+(fr.b)=(fp.1)*(fr1.1)+(fr.1));
\end{verbatim}}

In this version no attributes are used. The condition that the $m_i$
are pairwise relatively prime is here stated explicitly using
the {\tt gcd} functor for natural numbers.
Also the congruences are described arithmetically: $u \equiv u_i$ mod $m_i$
means that there exists a $x_i$ such that $u = u_i + x_i * m_i$,
so the theorem basically states the existence of $x_1, \ldots , x_r$ instead
of $u$.

Since the article has been written more than 10 years ago, a reason for this
technical formulation is hard to find.
It may be that at the
time of writing Mizar's attribute mechanism was not so far developed as today, i.e.
the author reformulated the theorem in order to get it formalized at all.
Another explanation for this second technical version might be that
the author when formalizing the CRT
already had in mind a particular application
and therefore chose a formulation better suited to prove
the application.

In the Coq Proof Assistant \cite{coqhome} the CRT has been proved
for a bit vector representation of the integers \cite{coq},
though as a restricted version of Theorem \ref{cong} 
with two moduli {\tt a} and {\tt b}.

{\small\begin{verbatim}
Theorem chinese_remaindering_theorem :
 forall a b x y : Z,
 gcdZ a b = 1%Z -> {z : Z | congruentZ z x a /\ congruentZ z y b}.
\end{verbatim}}

\noindent
In fact this theorem and its proof are the result of rewriting a former proof of the CRT in Coq.
So in Coq there exist two versions of the CRT --- though the former one has been
declared obsolete.

We see that in general the way authors use open systems to formalize theorems has
a crucial impact on the formulation, that is on the technical version
of a theorem, and may lead to different versions of the same theorem.
Removing one --- usually the older --- version is a dangerous task:
In large repositories it is not clear whether all proofs relying on
the deleted version can be easily changed to work with the other one.
So often both versions reamin in the repository.

\section{Abstract and Concrete Mathematics} 

Practically every mathematical repository has a notion of groups, rings, fields
and many more abstract structures.
The advantage is obvious: A theorem shown to hold in an abstract structure is also true
in every concrete structure of this type.
This can help to kepp a repository small: Even if concrete structures are defined
there is no need to repeat theorems following from the abstract structure.
If necessary in a proof for a concrete structure one can just use the theorem
proved for the abstract structure.

Nevertheless authors tend to prove theorems again for the concrete case.
We can observe this phenomenon in the Mizar Mathematical Library (MML).
There we find, for example, the following theorem about groups.

{\small\begin{verbatim}
theorem
for V being Group 
for v being Element of V holds v - v = 0.V;
\end{verbatim}}

For a number of conrete groups (rings or fields) this theroem, however,
has been proved and stored in MML again, among them complex numbers
and polynomials.

{\small\begin{verbatim}
theorem
for a being complex number holds a - a = 0;

theorem
for L be add-associative right_zeroed right_complementable
        (non empty addLoopStr) 
for p be Polynomial of L holds p - p = 0_.(L);
\end{verbatim}}

One reason might be that authors are not aware of the abstract theorems
they can use and therefore think that it is necessary to include theorems
in the concrete case.
This might be especially true, if authors work on applications rather than
on "core" mathematics.
On the other hand it might just be more comfortable for authors to work solely
in the concrete structure rather than to switch between concrete and abstract
structures while proving theorems in a concrete structure.

Constructing new structures from already existent ones sometimes causes
a similar problem: Shall we formalize a more concrete or a more abstract
construction?
Multivariate polynomials, for example, can be recursively constructed from univariate polynomials
using $R[X,Y] \cong (R[X])[Y]$;
or more concrete as functions from Terms in $X$ and $Y$ into the ring $R$.
Which version is better suited for mathematical repositories?
Hard to say, from a mathematical point of view the first version is the more interesting
construction. The second one, however, seems more intuitive and may be more convinient
to apply in other areas where polynomials are used.
So, it might be reasonable to include both constructions in a repository.
In this case, however, theorems about polynomials will duplicate also.

We close this section with another example: rational functions.
Rational functions can be constructed as pairs of polynomials or as
the completion $K(X)$ of the polynomial ring $K[X]$.
As in the case of multivariate polynomials both constructions have its right in
its own, so again both may be included in a repository.
Note that this eventually might result in another (two) concrete version(s) of the theorem
about groups from above, e.g.

{\small\begin{verbatim}
theorem
for L being Field
for z being Rational_Function of L
holds z - [0_.(L),1_.(L)] = z;
\end{verbatim}}

\section{Representational Issues}

In the majority of cases it does not play a major role how mathematical objects
are represented in repositories.
Whether the real numbers, for example, are introduced axiomatically or
are constructed as the Dedekind-completion of the rational numbers,
has actually no influence on later formalizations using real numbers.
Another example are ordered pairs: Here we can apply Kuratowski's
or Wiener's definition that is
\[ (a,b) = \{\{a\}\},\ \{a,b\}\}\]
or
\[ (a,b) = \{\{\{a\},\emptyset\},\{\{b\}\}\} \]
or even again the axiomatically approach
\[ (a_1,b_1) = (a_2,b_2) \mbox{ if and only if } a_1 = a_2 \mbox{ and } b_1 = b_2.\]
Once there is one of the notions included in a repository
formalizations relying on this notion can be carried out more or less the same.

There are, however, mathematical objects having more  than one interesting
representation.
The most prominent example are polynomials. Polynomials can be 
straightforwardly constructed as
sequences (of coefficients) over a ring
\[ p = (a_n, a_{n-1}, \ldots a_0) \]
or as functions from the natural numbers into a ring
\[ p = f: I\!\!N \longrightarrow R \mbox{ where } |\{ x | f(x) \neq 0 \}| < \infty. \]
Note that both representations explicitely mention all zero coefficients of a polynomial,
that is provide a dense representaion.

There is an alternative seldom used in repositories: sparse polynomials.
In this representation only coefficients not equal to $0$ are taken into account --- at
the cost that exponents have to be attached. We thus get a list of pairs:
\[ p = ((e_1,a_1), (e_2,a_2), \ldots (e_m,a_m)) .\]
Though more technically to deal with --- that probably being the reason for
usually choosing a dense representation for formalization --- there exist 
a number of efficient algorithms based on a sparse representation, for example
interpolation and computation of integer roots.
Therefore it seems reasonable to formalize both representations in a repository,
thus reflecting the mathematical treatment of polynomials.

Another example is the representation matrices, also a rather basic mathematical structure.
The point here is that there exist many interesting subclasses of matrices,
for example block matrices for which a particular multiplication algorithm can be given or
triangular matrices for which equations are much easier to solve.
Hence it might be reasonable to include different representations of matrices,
that is different (re-) definitions,
in a repository to provide support for particular applications of matrices.

\section{Generalization of Theorems}

Generalization of theorems is everyday occurrence in mathematics.
In the case of mathematical repositories generalization is a rather
involved topic: It is not obvious whether the less general
theorem can be eliminated. Proofs of other theorems using the original
version might not work automatically with the more general theorem instead.
The reason may be that a slightly different formulation or even a different 
(mathematical or technical) version of the original theorem has been formalized.
Then the question is: Should one rework all these proofs or keep both the original
and the more general theorem in the repository?
To illustrate that this decision is both not trivial and important
for the organization of mathematical repositories we present
in this section some generalizations of the CRT.

A rather harmless generalization of Theorem \ref{cong} is based on the observation
that the range in which the integer $u$ lies, does not need to be fixed.
It is sufficient that it has the width $m = m_1m_2\cdots m_r$.
This easily follows from the properties of the congruence $\equiv$.

\begin{theorem} \rm \label{g1}
Let $m_1, \, m_2, \, \ldots , m_r$ be positive integers such that $m_i$ and $m_j$
are relatively prime for $i \neq j$.
Let $m = m_1 \, m_2 \, \cdots m_r$ and let $a, u_1, \, u_2, \, \ldots , u_r$
be integers.
Then there exists exactly one integer $u$ with 
\[ a \leq u < a + m \mbox{ and } u \ \equiv \ u_i \mbox{ mod } m_i \]
for all $1 \leq i \leq r$. $ \ \diamond $ 
\end{theorem}

It is trivial that for $a = 0$ we get the original Theorem \ref{cong}.
Old proofs can very easily be adapted to work with this generalization
of the theorem.
Maybe the system checking the repository even automatically infers that 
Theorem \ref{g1} with $a = 0$ substitutes the original theorem.
If not, however, even the easy changing all the proofs to work with the
generalization can be an extensive, unpleasant, and time-consuming task.

A second generalization of the CRT is concerned with
the underlying algebraic structure.
The integers are the prototype example for Euclidean domains.
Taking into account that the residue class ring $\mathcal{Z}_n$ in fact is
the factor ring of $\mathcal{Z}$ by the ideal $n\mathcal{Z}$,
it is rather obvious that the following generalization\footnote{Literally
this is a generalization of Theorem \ref{alg}, but of course Theorem
\ref{cong} can be analogously generalized to Euclidean domains.} holds.

\begin{theorem} \rm \label{g2}
Let $R$ be a Euclidean domain.
Let $m_1, \, m_2, \, \ldots , m_r$ be positive integers such that $m_i$ and $m_j$
are relatively prime for $i \neq j$ and
let $m = m_1 \, m_2 \, \cdots m_r$.
Then we have the ring isomorphism
\[ R / (m) \ \cong \  R / (m_0) \times \cdots \times  R / (m_r). \ \ \diamond \]
\end{theorem}

This generalization may cause problems: In mathematical repositories
it is an important difference whether one argues about the set of integers (with
the usual operations) or the ring of integers:
They have just different types.
Technically, this means that in mathematical repositories we 
often have two different representations of the integers.
In the mathematical setting theorems of course hold for both of them.
However, proofs using one representation will not automatically work for the other one.
Consequently, though Theorem \ref{g2} is more general, it will not work
for proofs using integers instead of the ring of integers; 
for that a similar generalization of Theorem \ref{cong} is necessary.
So in this case in order to make all proofs work with a generalization,
we need to provide generalizations of different versions of the original theorem
--- or just change the proofs with the ``right'' representation
leading to an unbalanced organization of the repository.

We close this subsection with a generalization of the CRT
that abstracts even from algebraic structures.
The following theorem \cite{luneburg} deals with sets and equivalence relations
only
and presents a condition whether the ``canonical'' function  $\sigma$ is onto.

\begin{theorem} \rm \label{sets}
Let $\alpha$ and $\beta$ be equivalence relations on a given set $M$.
Let $\sigma: M \longrightarrow M/\alpha \times M/\beta$ be defined
by $\sigma(x) := (\alpha(x), \beta(x))$.
Then 
we have $\mbox{ker}(\sigma) = \alpha \cap \beta$ and
$\sigma$ is onto if and only if $\alpha \circ \beta = M \times M$.
 $ \ \diamond $ 
\end{theorem}

Here almost all of the familiar CRT
gets lost. There are no congruences, no algebraic operations, only the
factoring (of sets) remains.
Therefore, it seems hardly possible to adapt proofs using any of the preceding
CRTs to work with this generalization in a reasonable
amount of time.
Any application will rely on much more concrete structures, so that too much effort has to be spent
to adapt a proof.
Theorem \ref{sets} in some sense is too general to reasonably work with.
However, even though hardly applicable, the theorem stays interesting from a didactic
point of view.\footnote{In fact the proof of Theorem \ref{sets} has been an exercise in 
lectures on linear algebra.}
It illustrates how far we sometimes can generalize and may provide the starting point of
a discussion whether this is --- aside from mathematical aesthetics ---
expedient;
a topic that is also of great interest for the organization of mathematical repositories.

\section{Conclusions}

When building a mathematical repository it seems plausible to not duplicate knowledge
in order to avoid an unnecessary blow-up of the repository.
This is similar to --- and may be inspired by --- mathematical definitions, in which
the number of axioms is kept as small as possible.

In this paper we have argued that this, however, is not true in general.
We have analyzed miscellanous situations in which it might be reasonable or even necessary
to duplicate knowledge in a repository.
The reasons for that are manifold:
Different proofs may be interesting for didactic reasons or
different representations of the same knowledge may better support different groups of users.
Even improvements of a repository may lead to duplication of knowledge because e.g.
the improved version of a theorem cannot always be trivially erased without
reworking lots of proofs.

In general, it is hardly foreseeable in which cases which kind of knowledge should be duplicated.
This strongly depends on different kind of users the repository should attract.
 \vspace*{0,3cm}\\


\begin{thebibliography}{ABCD99}


\bibitem[Coq10]{coqhome}
  {\em The Coq Proof Assistant};
  available at {\tt http://coq.inria.fr}.
 
\bibitem[Dav03]{Davenport} 
  J.~H.~Davenport,
  {\em MKM from Book to Computer: A Case Study};
  in: A. Asperti, B. Buchberger, and J. Davenport (eds.), Proc. of MKM 2003,
  Lecture Notes in Computer Science 2594, pp. 17-29, 2003.


\bibitem[DeB87]{DeBruijn} 
  N.G. de~Bruijn,
  {\em The Mathematical Vernacular, a language for mathematics
  with typed sets}; in: P. Dybjer et al. (eds.), Proceedings of the Workshop
  on Programming Languages, Marstrand, Sweden, 1987.


\bibitem[GG99]{modern}
  J. von zur Gathen and J. Gerhard, 
  {\em Modern Computer Algebra};
  Cambridge University Press, 1999.

\bibitem[GS06]{vernacular}
  A. Grabowski and C. Schwarzweller,
  {\em Translating Mathematical Vernacular into Knowledge Repositories};
  in: M. Kohlhase (ed.), Proceedings of the 4th International Conference
  on Mathematical Knowledge Management,
  Lecture Notes in Artificial Intelligence 3863, pp. 49-64, 2006.



\bibitem[Har10]{hollight}
  J. Harrison,
  {\em The HOL Light Theorem Prover};
  available at {\tt http://} \verb@www.cl.cam.ac.uk/~jrh13/hol-light@.

\bibitem[Hur03]{hurd}
  J. Hurd,
  {\em Verification of the Miller-Rabin Probabilistic Primality Test};
  in: Journal of Logic and Algebraic Programming, 50(1-2), pp. 3-21, 2003.

\bibitem[KZ89]{rrl}
  D. Kapur and H. Zhang,
  {\em An Overview of Rewrite Rule Laboratory (RRL)};
  in: N. Dershowitz (ed.), Proceedings of the 3rd International Conference on Rewriting Techniques 
  and Applications, LEcture Notes in Computer Science 355, pp. 559-563, 1989.

\bibitem[KN04]{Fairouz}
  F.~Kamareddine and R.~Nederpelt,
  {\em A Refinement of de Bruijn's Formal Language of Mathematics}; in:
  in: Journal of Logic, Language and Information, 13(3), pp. 287-340, 2004.

\bibitem[Knu97]{knuth}
  D. Knuth, 
  {\em The Art of Computer Programming, Vol. 2: Seminumerical Algorithms};
  3rd edition, Addison-Wesley, 1997.

\bibitem[Kon97]{wsierp1}
  A. Kondracki,
  {\em The Chinese Remainder Theorem};
  in: Journal of Formalized Mathematics, vol. 6(4), pp. 573-577, 1997.

\bibitem[L\"{u}n93]{luneburg}
  H. L\"{u}neburg, 
  {\em Vorlesungen \"{u}ber Lineare Algebra}; (in German),
  BI Wissenschaftsverlag, 1993.

\bibitem[M\'{e}n10]{coq}
  Val\'{e}rie M\'{e}nissier-Morain,
  {\em A Proof of the Chinese Remainder Lemma};
  available at 
  {\tt http://logical.saclay.inria.fr/coq/distrib/current/} {\tt contribs/ZChinese.html}.

\bibitem[Miz10]{HomePage}
  The Mizar Home Page, {\tt http://mizar.org}.

\bibitem[NB04]{Naumowicz}
  A. Naumowicz and C. Byli\'nski, 
  {\em Improving Mizar Texts with Properties and Requirements},
  in: A.~Asperti, G. Bancerek, and A. Trybulec (eds.),
  Proceedings of the 3rd International Conference on Mathematical Knowledge Management,
  Lecture Notes in Computer Science 3119, pp. 190-301, 2004.

\bibitem[PSK04]{matrices}
  Martin Pollet, Volker Sorge, and Manfred Kerber,
  {\em Intuitive and Formal Representations: The Case of Matrices};
  in: A.~Asperti, G. Bancerek, and A. Trybulec (eds.),
  Proceedings of the 3rd International Conference on Mathematical Knowledge Management,
  Lecture Notes in Computer Science 3119, pp. 317-331, 2004

\bibitem[RT01]{mizar}
  P. Rudnicki and A. Trybulec, 
  {\em Mathematical Knowledge Management in Mizar};
  in: B. Buchberger, O. Caprotti (eds.), Proceedings of the 1st International 
  Conference on Mathematical Knowledge Management,
  Linz, Austria, 2001.

\bibitem[Sch08]{int6}
  C. Schwarzweller, 
  {\em Modular Integr Arithmetic};
  in: Journal of Formalized Mathematics, vol. 16(3), pp. 247-252, 2008.

\bibitem[Sch09]{slgr09}
  C. Schwarzweller,
  {\em The Chinese Remainder Theorem, its Proofs and its Generalizations in Mathematical Repositories};
  in: Studies in Logic, Grammar and Rhetoric 18(31), p. 103-119, 2009. 

\bibitem[Sho67]{thint}
  J. Shoenfield,
  {\em Mathematical Logic};
  Addison-Wesley, 1967.

\bibitem[Wie06]{useful}
  F. Wiedijk, 
  {\em On the Usefulness of Formal Methods};
  Nieuwsbrief van de NVTI, pp. 14-23, 2006.

\bibitem[ZH92]{zhang}
  H. Zhang and X. Hua,
  {\em Proving the Chinese Remainder Theorem by the Cover Set Induction};
  in: D. Kapur (ed.), Automated Deduction - CADE-11, Proceedings of the
  11th International Conference on Automated Deduction,
  Lecture Notes in Computer Science 607, pp. 431-455, 1992. 

\end{thebibliography}
\end{document}